\documentclass[useAMS,usenatbib]{mn2e}
\usepackage{mncite}
\usepackage{graphicx}
\usepackage{epstopdf}
\usepackage{subfigure}
\usepackage[a4paper]{hyperref}
\usepackage{amssymb}
\usepackage{aas_macros}
\usepackage[usenames]{color}

\newcommand{\rmsub}[2]{#1_{\rm #2}} 

\title[Stellar rotation in Coma Berenices]{The main-sequence rotation-colour relation in the Coma Berenices open cluster}
\author[A. Collier Cameron et al]
{
A. Collier Cameron$^{1}$\thanks{E-mail:acc4@st-and.ac.uk},
V. A. Davidson$^{1}$,
L. Hebb$^{1}$,
G. Skinner$^{1}$,
D. R. Anderson$^{2}$,
\newauthor
D. J. Christian$^{3}$,
W. I. Clarkson$^{4}$,
B. Enoch$^{1}$,
J. Irwin$^{5}$,
Y. Joshi$^{3}$,
C. A. Haswell$^{6}$,
\newauthor
C. Hellier$^{2}$,
K. D. Horne$^{1}$,
S. R. Kane$^{7}$,
T. A. Lister$^{8}$,
P. F. L. Maxted$^{2}$,
\newauthor
A. J. Norton$^{6}$,
N. Parley$^{6}$,
D. Pollacco$^{3}$,
R. Ryans$^{3}$,
A. Scholz$^{1}$,
I. Skillen$^{9}$ ,
\newauthor
B. Smalley$^{2}$,
R. A. Street$^{8}$,
R. G. West$^{10}$,
D. M. Wilson$^{2}$
and
P.J. Wheatley $^{11}$
\\
$^{1}$SUPA, School of Physics and Astronomy, University of St Andrews, North Haugh, St Andrews, Fife KY16 9SS, UK.\\
$^{2}$Astrophysics Group, School of Chemistry and Physics, Keele University, Staffordshire, ST5 5BG, UK.\\
$^{3}$Astrophysics Research Centre, School of Mathematics \&\ Physics, Queen's University, University Road, Belfast, BT7 1NN, UK.\\
$^{4}$STScI, 3700 San Martin Drive, Baltimore, MD 21218, USA.\\
$^{5}$Harvard-Smithsonian Center for Astrophysics, 60 Garden St, Cambridge, MA, USA.\\
$^{6}$ Department of Physics and Astronomy, The Open University, Milton Keynes, MK7 6AA, UK.\\
$^{7}$Michelson Science Center, Caltech, MS 100-22, 770 South Wilson Avenue, Pasadena, CA  91125, USA.\\
$^{8}$Las Cumbres Observatory, 6740 Cortona Dr. Suite 102,
Santa Barbara, CA 93117, USA.\\
$^{9}$Isaac Newton Group of Telescopes, Apartado de Correos 321, E-38700 Santa Cruz de la Palma, Tenerife, Spain. \\
$^{10}$Department of Physics and Astronomy, University of Leicester, Leicester, LE1 7RH, UK.\\
$^{11}$Department of Physics, University of Warwick, Coventry CV4 7AL, UK.
}
\begin{document}

\date{Accepted 2009 July 30.  Received 2009 July 30; in original form 2009 June 3}

\pagerange{\pageref{firstpage}--\pageref{lastpage}} \pubyear{2009}

\maketitle

\label{firstpage}

\begin{abstract}
We present the results of a photometric survey of rotation rates in the Coma Berenices (Melotte 111) open cluster, using data obtained as part of the SuperWASP exoplanetary transit-search programme. The goal of the Coma survey was to measure  precise rotation periods for main-sequence F, G and K dwarfs in this intermediate-age ($\sim$ 600 Myr) cluster, and to  determine the extent to which magnetic braking has 
caused the stellar spin periods to converge. We find a tight, almost linear relationship between rotation period and $J-K$ colour with a root-mean square scatter of only 2 percent.  The relation is similar to that seen among F, G and K stars in the Hyades. Such strong convergence can only be explained if angular momentum is not at present being transferred from a reservoir in the deep stellar interiors to the surface layers.  We conclude that the coupling timescale for angular momentum transport from a rapidly-spinning radiative core to the outer convective zone must be substantially shorter than the cluster age, and that from the age of Coma onward, stars rotate effectively as solid bodies. The existence of a tight relationship between stellar mass and rotation period at a given age supports the use of stellar rotation period as an age indicator in F, G and K stars of Hyades age and older.  We demonstrate that individual stellar ages can be determined within the Coma population with an internal precision of order 9 percent (RMS), using a standard magnetic braking law in which rotation period increases with the square root of stellar age. We find that a slight modification to the magnetic-braking power law, $P\propto t^{0.56}$, yields rotational and asteroseismological ages in good agreement for the Sun and other stars of solar age for which $p$-mode studies and photometric rotation periods have been published.
\end{abstract}


\section{Introduction}

In their pioneering study of stellar rotation among main-sequence stars in the Hyades open cluster, \citet{radick87hyades} found a surprisingly narrow correlation between the $B-V$ colours and the rotation periods  of F, G and K stars. This discovery has several far-reaching implications. First, it confirms the theoretical prediction that the spin rates of an ensemble of stars of the same age and mass, but different initial rotation periods rates, should converge. This comes about because the rate at which angular momentum is lost through magnetic braking via a thermally-driven, magnetically-channelled wind, increases strongly as a function of stellar rotation rate (e.g. \citealt{mestel87braking}). Radick et al showed that this convergence is essentially complete at the $\sim 600$ Myr age of the Hyades. Second, for convergence to occur at such a comparatively early age, the timescale for coupling the star's radiative interior to its outer convective zone must be short. Models that allow a star's outer convective zone to be spun  down rapidly by wind losses at the 50-100 Myr ages of clusters such as the Pleiades, retaining a hidden reservoir of angular momentum in a  rapidly-spinning radiative interior, do not show this convergence 
(\citealt{li93rot}; \citealt{krishnamurthi97rot}; \citealt{allain98rot}). 

For the core and envelope to be decoupled at the age of the Pleiades but to be rotating synchronously by the age of the Sun, the coupling timescale must be no less than $\sim 10$ Myr, and no more than 1 Gyr \citep{soderblom2001m34}. Rotational evolutionary tracks with coupling timescales of order 100 Myr do not converge, because the ``buried" angular momentum re-emerges at ages of a few hundred Myr. The wide range of spin rates seen in the very youngest clusters then re-asserts itself, inducing a large scatter in the predicted rotation rates and delaying convergence well beyond the age of Coma and the Hyades. 

The Hyades study by \citet{radick87hyades} remained unparallelled for nearly two decades after its publication. The advent of wide-field synoptic surveys for exoplanetary transits in open clusters using 1m-class telescopes, and in nearby field stars using even smaller instruments, has led to a recent resurgence of interest in the rotation distributions of open clusters with ages of $\sim 150$ Myr and older. Several recent photometric studies show the convergence of spin rates in nearby open clusters of intermediate age, notably M35 (150 Myr; \citealt{meibom2009m35}), M34 (250 Myr: \citealt{irwin2006m34}) and M37 (550 Myr: \citealt{hartman2008m37}; \citealt{messina2008m37}). These studies have gained new significance because the rotation-age-colour relations in these clusters provide important calibration points for stellar rotational evolution. An empirical calibration of stellar spin rates as a function of  age and colour was published recently by \citet{barnes2007gyro},
\textcolor{black}{
who coined the term {\it gyrochronology} to describe the inverse process of inferring a star's age from its rotation period and colour.
}
If magnetic braking causes stellar spin rates to converge by the age of the Hyades (or perhaps sooner) and subsequently to spin down  as some unique and measurable function of time, gyrochronology becomes very attractive as a clock for measuring the ages of individual field stars, stellar and planetary systems.  

While this convergence of spin rates seems to be essentially complete among F and G stars by the age of M35 \citep{meibom2009m35}, and in K stars by the age of the Hyades, it takes even longer for stars of still lower mass. Although many stars with $B-V>1.3$ have converged to the main period-colour relation by the 550-Myr age of M37, \citet{hartman2008m37} identified a substantial clump of stars redder than this limit having periods of 2.5 days or less. In their preliminary investigation of the still older Praesepe (650 Myr), \citet{scholz2007praesepe} find convergence to be complete down to the late K types, but rapid rotation persisting among low-mass stars of spectral type M0 and later.

Here we present the results of a wide-field photometric monitoring survey of
stars brighter than $V=14$ in the region around the Coma Berenices open cluster,
using the SuperWASP camera array on La Palma. 
This cluster, also known as Melotte 111, is centred at RA = 12h 23m,  Dec. = +26$^\circ$ 00Õ (J2000.0). It is the second-closest open cluster to the Sun, lying at  a
distance of $89.9\pm 2.1$ pc \citep{vanleeuwen99clus}, further only than the 
Hyades (46.3 pc: \citealt{perryman97}). The foreground reddening is 
$E(B-V) \simeq 0.006 \pm 0.013$ \citep{nicolet81coma}. \citet{trumpler38coma} 
\textcolor{black}{
used proper-motion, radial-velocity and colour-magnitude selection criteria to identify
}
37 probable members with photographic magnitudes $m_{pg}<10.5$ within 3.5$^\circ$ of the cluster centre. Fainter candidates with $m_{pg}<15.0$ were identified within the same region in a  proper motion survey by \citet{artiukhina55coma}, but only two of these were
confirmed by  \citet{argue69coma}, who identified another two faint
objects that were later  confirmed as members.  \citet{odenkirchen98coma}
performed a thorough kinematic and  photometric survey down to $V\simeq 10.5$
using the {\em Hipparcos} and 
\textcolor{black}
{the combined Astrographic Catalogue and {\em Tycho} (ACT; \citealt{urban98act})  }
reference catalogues. This study yielded
around 50 probable kinematic members in a  region of $1200$ square degrees
about the cluster centre, 
\textcolor{black}
{effectively superseding the kinematic and photometric aspects of Trumpler's original survey.}
\citet{casewell2006coma} compiled a list of previously known members 
\textcolor{black}
{(according to proper-motion, photometric and in some cases radial-velocity criteria) within 4 degrees of the cluster centre. They
found a further 60 candidate cluster members using} the USNO-B1.0
(United States Naval Observatory; \citealt{monet2003}) and 2MASS (Two-Micron All-Sky Survey)
point-source catalogues to carry out proper motion and photometric surveys
respectively, more than doubling the number of probable Melotte 111 cluster
members 
\textcolor{black}
{to about 110 according to photometric and kinematic criteria. Among these, about 40 
are secure in the sense that they also satisfy radial-velocity membership criteria, at 
least at the level of precision achieved by \citet{trumpler38coma}.}

Previous photometric studies by \citet{radick90coma} of four early-G dwarfs and by \citet{marilli97coma} 
of three late-G dwarfs in the Coma Berenices cluster suggest a Hyades-like convergence in spin rates. Like the Hyades, however, 
Coma Berenices is a difficult target for high-precision photometric rotation studies, 
because of its age and and its wide angular extent. As in
the Hyades, the stellar spin periods are expected to range from 6 days among late F dwarfs to 
14 days among mid-K dwarfs, with more rapid rotation persisting at later spectral types. Although
stellar dynamos still produce strong magnetic activity signatures at such
rotation rates, the starspot coverage is substantially lower than in the
ultra-fast rotators found in younger open clusters. The comparatively low starspot coverage 
on stars in the Hyades and Coma yields amplitudes of rotational modulation of at most one or two percent. This combination of long periods, low amplitudes and 
large angular separation between cluster members presents a challenge to observers seeking precise measurements of their optical modulation periods.

In sections 2 and 3 of this paper we describe the observations and data reduction
methodology. In section 4 we describe the generalised Lomb-Scargle period-search method we used to identify candidate rotational variable stars in the Coma Berenices region, while in section 5 we describe the 
additional photometric and astrometric criteria used to assess their cluster membership probabilities. In Sections 6 and 7 we discuss the period-colour relation for Coma Berenices 
cluster members, and the implications for stellar spindown and gyrochronological age determination.

\section{Observations}

The SuperWASP camera array, located at the Observatorio del Roque de los
Muchachos on La Palma, Canary Islands, carried out its inaugural season of 
observations during 2004. The camera
array is mounted on a robotic equatorial mount. In 2004 it  comprised five 200mm f/1.8 Canon
lenses each with an Andor CCD array of $2048^2$ 13.5$\mu$m pixels, giving a field of
view 7.8 degrees square for each camera \citep{pollacco2006wasp}. 
Four individual camera fields covering the region around the Coma Cluster were  observed on
51 nights of good photometric quality between 2004 May 2 and 2004 July 18; a further two fields
at an adjacent pointing position were observed on 86 nights between 2004 May 02 and 2004 
August 12. 

The Coma Berenices region was re-observed throughout the first half of 2007. By this time a 
further three cameras had been installed on the mount. Again, six individual camera fields 
provided full coverage of the region around the cluster: two on 72 nights from 2006 December 
12 to 2007 May 15, and four from 2007 January 14 to May 30.

The survey region was accessible for between  4 and 8 hours each night. The
average interval between visits to the field was 6 minutes. Each exposure was
of 30s duration, and was taken without filters in 2004, and with an infrared blocking filter in 2007. 

\begin{table}
\caption{Journal of SuperWASP observations for the Coma Berenices rotation survey.}
\label{tab:journal}
\begin{tabular}{cccrr}
\hline\\
Field centre	& No of	& Usable	& Start	& Baseline \\
hhmm+ddmm	& images	& nights	& date	& (nights) \\
\hline\\
1143+3126	& 1200	& 51	& 2/5/04	& 78 \\
1144+2427	& 1203	& 52	& 2/5/04	& 78 \\
1216+3126	& 1111	& 51	& 2/5/04	& 78 \\
1217+2326	& 1200	& 51	& 2/5/04	& 78 \\
1243+3126	& 2378	& 86	& 2/5/04	& 103 \\
1244+2427	& 2383	& 86	& 2/5/04	& 103 \\

1217+2350	& 4876	& 71	& 30/12/06	& 137 \\
1222+3000	& 4926	& 72	& 30/12/06		& 137 \\
1238+3135	& 5815	& 78	& 14/1/07	& 137 \\
1241+3924	& 5796	& 77	& 14/1/07	& 137 \\
1242+2418	& 5667	& 76	& 14/1/07	& 137 \\
1252+1735	& 5707	& 77	& 14/1/07	& 137 \\
\hline\\
\end{tabular}
\end{table}

The field centres and dates of observation are summarised in Table~\ref{tab:journal}.

\section{Data reduction and calibration}

The data were reduced using the SuperWASP 
\textcolor{black}{data reduction pipeline, whose operation is described in detail by \citep{pollacco2006wasp}.} 
Science frames are bias-subtracted and
flat-fielded.  An automated field recognition algorithm identifies the objects
on the frame with their counterparts in the {\sc tycho-2} catalogue and
establishes an astrometric solution  with an RMS precision of 0.1 to 0.2 pixel.
\textcolor{black}{The pipeline then performs aperture photometry} at  the positions on each CCD image of
all objects in the USNO-B1.0 catalogue \citep{monet2003} with magnitudes in 
the $R2$ band brighter than 14.5, equivalent to $V\simeq 15$ over the colour range $0<B-V<1$. 
\textcolor{black}{The pipeline computes he formal variance of the flux of each star 
in the image} from the photon counts in the stellar aperture and the 
surrounding sky annulus. This magnitude limit is dictated by the detection 
threshold for the SuperWASP exoplanetary transit-search project; stars
with $R2\simeq 14.5$ yield light curves with RMS scatter between 0.02 and 
0.03 magnitude.

The SuperWASP bandpass covers most of the optical spectrum, so the effective
atmospheric extinction depends on stellar colour. The extinction correction to
zero airmass for a star of known $B-V$ colour index takes the form
$$
m_0 = m - k' X - k''(B-V) X
$$
where $m$  is the raw instrumental magnitude at airmass $X$, $m_0$ is the instrumental magnitude above the atmosphere, and $k'$ and $k''$ are the primary and secondary extinction coefficients 
respectively. The zero-point $z$ for each image is tied to a network of local secondary standards in each field. The
resulting instrumental magnitudes are transformed to {\sc tycho-2} $V$ magnitudes via the
colour equation
$$
V_T = m_0 - z - c (B-V) 
$$
where the colour transformation coefficient for all cameras lies in the range $c=-0.22\pm 0.03$. 
The resulting fluxes are stored in the SuperWASP Data Archive at the University of Leicester. 

We extracted the light-curves of all stars observed in the six fields 
nearest to the centre of  the Coma cluster in each of the 2004 and 2007 
observing seasons from the archive. We stored them in a two-dimensional  array whose columns held the light curves of individual stars, and whose rows held data derived from individual CCD frames. Patterns of correlated systematic 
error were identified and removed using the  {\sc SysRem}  algorithm of \citet{tamuz2005}, as implemented by \citet{cameron2006methods}. 
At this stage the photometric variances are augmented by an additional systematic variance that accounts for the actual scatter in the measured fluxes about the final decorrelated solution for each frame. The maximum-likelihood algorithm for computing this additional variance is described by \citet{cameron2006methods}.
The long-term RMS scatter of the decorrelated SuperWASP light curve of a non-variable star is 
0.004 mag at V = 9.5, degrading to 0.01 mag at V = 12.0 as shown in Fig. 3 of
\citet{cameron2006methods}.

\section{Light curve analysis}

\subsection{Frequency analysis}

We searched for evidence of quasi-sinusoidal light-curve modulation in all stars in the field, using the generalised Lomb-Scargle periodogram formulation of \citet{zechmeister2009} to fit an inverse variance-weighted floating-mean sinusoid to the light curve over a full observing season.

Each star's light curve comprises a set of observed magnitudes $m'_i\in\{m'_1, m'_2, ..., m'_N\}$ with associated variances $\sigma^2_i\in\{\sigma^2_1,\sigma^2_2,...,\sigma^2_N\}$. We
form differential magnitudes 
$$
m_i = m'_i - \hat{m},
$$
where
$$
\hat{m}=\frac{\sum_i m'_i/\sigma^2_i}{\sum_i 1/\sigma^2_i}.
$$
The goodness of fit of this constant-magnitude model to the light curve is given by the statistic
$$
\chi^2_0=\sum(m_i)^2/\sigma^2_i.
$$

We fit a sinusoidal model 
$$
y_i = A\cos\omega t_i + B\sin\omega t_i + C
$$
with angular frequency $\omega$ to the residuals $m_i$
at the times $t_i$ of observation. We solve for the coefficients $A$, $B$ and $C$  using the algorithms of \citet{zechmeister2009}. The light curve amplitude at frequency $\omega$ is then
$$
\delta m =\sqrt{\hat{A}^2+\hat{B}^2}
$$
and its associated variance is
$$
{\rm Var}(\delta m)=\frac{\hat{A}^2{\rm Var}(\hat{A})+\hat{B}^2{\rm Var}(\hat{B})}
{\hat{A}^2+\hat{B}^2}.
$$

We repeat the process at a set of frequencies corresponding to periods ranging from 1.11 days to 20 days.
\textcolor{black}
{At each frequency $\omega$ we compute the statistic
$$
p(\omega) = \frac{\chi^2_0-\chi^2(\omega)}{\chi^2_0}.
$$ 
For each star, the peak value $p_{\rm best}$ of the $p$ statistic is achieved at the
the frequency at which  $\chi^2$ is minimised.}

For a light curve comprising $N$ independent measurements, \citet{cumming99} estimate
\begin{equation}
{\rm Prob}(p>p_{\rm best})=(1-p_{\rm best})^{(N-3)/2}.
\label{eq:probp}
\end{equation}
The false-alarm probability 
\textcolor{black}
{(FAP)}
for $M$ independent frequency measurements is then
\begin{equation}
{\rm FAP} = 1 - [{\rm Prob}(p > p_{\rm best})]^M
\label{eq:fap}
\end{equation}
where $M\simeq T(f_2-f_1)$ for a data train of total duration $T$ searched in the frequency range $f_1 < f < f_2$ \citep{cumming2004}.

The false-alarm probability is sensitive to the number $N$ of independent observations. As \citet{cameron2006methods} found, some forms of correlated error repeat at intervals of one day, and are not adequately removed by the   {\sc SysRem} algorithm. A small misalignment of SuperWASP's polar axis causes every stellar image to drift across the frame by a few tens of pixels during each night. The vignetting pattern of the camera lenses produces discontinuities in the focal-plane illumination gradient near the edge of the frame, which change slightly from night to night and even during the course of a single night. The resulting correlated errors serve to reduce the number of independent observations from $N$ to a lower value $N_{\rm eff}$. 

We estimated $N_{\rm eff}$ by randomising the order of the list of dates on which data were obtained. The shuffling procedure shifts each night's observations in their entirety to a new date. This procedure effectively destroys coherent signals with periods longer than 1 day, but retains the global form of the window function and the effects of correlated noise. 
\textcolor{black}
{If many such trials are made, we expect $p_{\rm best}$ in their individual periodograms to exceed the value found for a single randomly-chosen trial roughly half the time. We thus estimate a false-alarm probablility FAP$_0=0.5$ for the the strongest peak $p_{\rm best}$ in the periodogram of the shuffled data. Using this estimate, we combine and invert equations \ref{eq:probp} and \ref{eq:fap} to}
obtain an estimate of the effective number of independent observations in the presence of correlated errors: 
\begin{equation}
N_{\rm eff} \simeq \frac{3 \ln(1-p_{\rm best})+2\ln(1-(1-FAP_0)^{1/M})}{\ln(1-p_{\rm best})}.
\label{eq:neff}
\end{equation}
In practice we found that typically $N_{\rm eff}\simeq N/10$. Since SuperWASP observes each field at intervals
of 8 minutes or so, the effective correlation time is of order 1.5 hours. 

\begin{figure}
\includegraphics[width=7.5cm,angle=270]{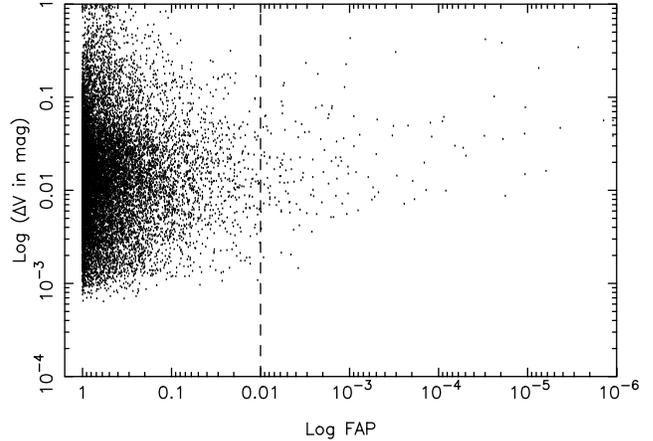}
\caption[]{Amplitude of best-fit sinusoidal model as a function of false-alarm probability for stars in the field SW1216+3126 in the 2004 season. Stars to the right of the selection threshold set at  FAP$ = 10^{-2}$ are considered to be candidate rotational variables.}
\label{fig:amp_fap}
\end{figure}
In Fig.~\ref{fig:amp_fap} we plot the  best-fitting modulation amplitude against false-alarm probability for all stars in the field SW1216+3126 observed in the 2004 season. We select for further study all those light curves showing evidence of periodic modulation with false-alarm probabilities less than $10^{-2}$. These stars exhibit light curve amplitudes in the range 0.01 to 0.1 mag, which is typical of rotational variables with periods of order a few days.


\section{Candidate selection}

Having identified a large number ($N_{\rm tot}=1613$) of rotational variables in the vicinity of the cluster, the next step is to determine which among them are likely to be cluster members. We followed the method of \citet{girard89clusmem}, defining probability density functions for the proper motions of $N_c$ cluster members and $N_f = N_{\rm tot}-N_c$ field stars. 

\begin{figure}
\includegraphics[width=7.5cm,angle=270]{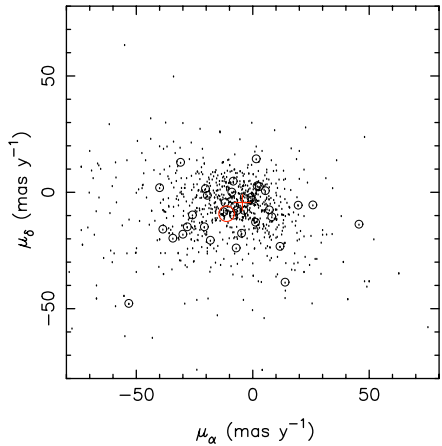}
\caption[]{Proper-motion diagram of stars with significant (FAP$ < 10^{-2}$) quasi-sinusoidal variability and periods between 1.1 and 20 days, in the WASP fields surveyed in 2004 and 2007. The 
\textcolor{black}
{red}
cross denotes the centroid of the field distribution for stars with total proper motions less than 25 mas/yr, while the 
\textcolor{black}
{red}
circle marks the mean proper motion of the cluster members. 
\textcolor{black}
{The 30 stars with combined kinematic and photometric membership probabilities greater than 0.5 (see also Table \ref{tab:starparms}) are encircled. }
}
\label{fig:propermotion}
\end{figure}

We obtained proper motions and magnitudes in the $V$ and 2MASS $J$ and $K_s$ bands for all candidate rotational variables by cross-matching with the NOMAD-1 catalogue \citep{zacharias2005nomad}. Following \citet{casewell2006coma} we adopt $\mu_\alpha=-11.21\pm 0.26$ mas yr$^{-1}$ and $\mu_\delta=-9.16\pm 0.15$ mas yr$^{-1}$ as the mean proper motion of the known cluster members, as determined by \citet{vanleeuwen99clus}. The proper-motion diagram for the full sample is shown in Fig.~\ref{fig:propermotion}. The field distribution for stars with total proper motions less than 25 mas y$^{-1}$ is approximated by a two-dimensional gaussian probability density function having the form
\begin{equation}
f_f = 
\frac{N_f}
{2\pi\Sigma_{f\alpha}\Sigma_{f\delta}}
\exp\left(-\frac{(\mu_{\alpha i}-\mu_{f\alpha})^2}{2\Sigma_{f\alpha}^2}-\frac{(\mu_{\delta i}-\mu_{f\delta})^2}{2\Sigma_{f\delta}^2}\right),
\label{eq:probdensf}
\end{equation}
with mean proper motion $\mu_{f\alpha}=-4.42$ mas yr$^{-1}$, $\mu_{f\delta}=-4.39$ mas yr$^{-1}$ and dispersion $\Sigma_{f\alpha}=9.90$ mas yr$^{-1}$, $\Sigma_{f\delta}=8.18$ mas yr$^{-1}$. 

\begin{figure}
\includegraphics[width=7.5cm,angle=270]{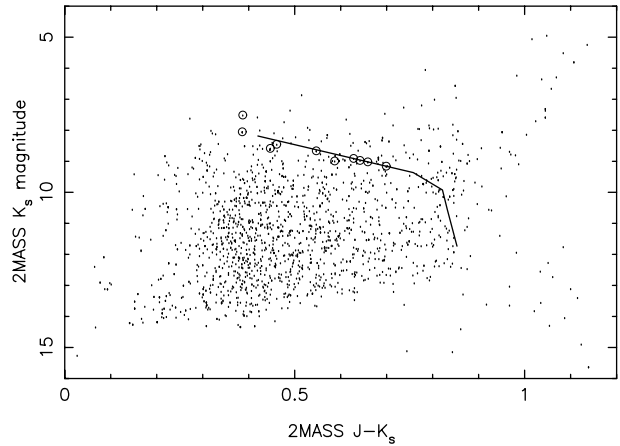}
\caption[]{2MASS ($K,J-K$) colour-magnitude diagram of stars with significant (FAP$ < 10^{-2}$) quasi-sinusoidal variability and periods between 1.1 and 20 days, in the WASP fields surveyed in 2004 and 2007. 
Previously-known cluster members detected as rotational variables and satisfying our cluster membership criteria are encircled.
The solid curve is the 520 Myr isochrone of \citet{baraffe98} for stars of solar metallicity at the distance of the Coma Berenices cluster. 
}
\label{fig:colmag}
\end{figure}

\begin{figure}
\includegraphics[width=7.5cm,angle=270]{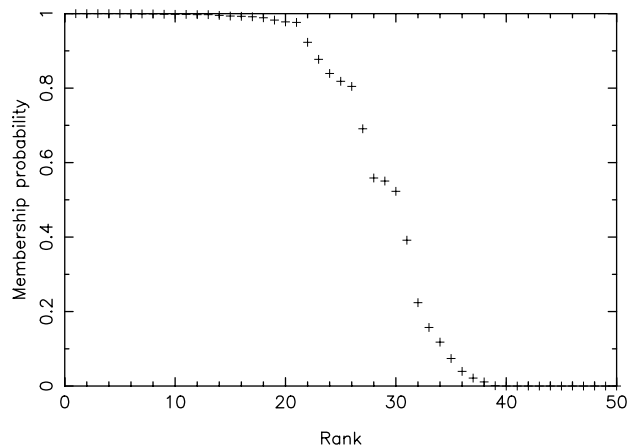}
\textcolor{black}
{
\caption[]{Ranked cluster membership probabilities of the top 50 stars (+) whose
light curves recorded by one or more WASP cameras in 2004 or 2007 showed significant (FAP$ < 10^{-2}$) quasi-sinusoidal variability and periods between 1.1 and 20 days. The membership probability is determined from the proper motions and K magnitude distributions for cluster and field stars, as described in the text. 30 stars have cluster membership probabilities greater than 0.5.
}
}
\label{fig:rank_pmem}
\end{figure}

For cluster members the density function depends primarily on the uncertainties in the proper-motion components $\sigma_{\alpha i}$ and $\sigma_{\delta i}$ for the individual stars, since the intrinsic spread in the proper motions of the cluster members is smaller than the measurement errors:
\begin{equation}
f_c = 
\frac{N_c}
{2\pi\sigma_{c\alpha}\sigma_{c\delta}}
\exp\left(-\frac{(\mu_{\alpha i}-\mu_{c\alpha})^2}{2\sigma_{\alpha i}^2}-\frac{(\mu_{\delta i}-\mu_{c\delta})^2}{2\sigma_{\delta i}^2}\right).
\label{eq:probdensc}
\end{equation}
Following \citet{girard89clusmem}, the cluster membership probability for an individual star is 
$$
p=\frac{f_c}{f_c+f_f}.
$$

We generalised this approach to take into account the magnitude distribution of field stars and proximity to the cluster main sequence in a colour-magnitude diagram constructed from the 2MASS $K$ magnitudes and $J-K$ colour index (Fig.~\ref{fig:colmag}). To select likely cluster members using their locations in this C-M diagram, we adopted the 520 Myr isochrone from Baraffe's (1998) NEXTGEN stellar evolution models, for low-mass stars of solar metallicity. The theoretical absolute $K$ magnitudes were converted to apparent magnitudes using the HIPPARCOS distance of $89.9\pm 2.1$ pc to the cluster \citep{vanleeuwen99clus}. The Baraffe $J-K$ colour and $K$ magnitude were converted from the CIT system for which they were computed to their 2MASS equivalents using the transformation of \citet{carpenter2001trans}. We allow an intrinsic scatter $\Sigma_{cK}= 0.15$ mag about the theoretical main sequence, assuming the depth of the cluster to be of order the 5 to 6 pc tidal radius estimated by \citet{odenkirchen98coma}. The probability density function is assumed to be gaussian.

Because the WASP bandpass is in the optical range, the colour-magnitude diagram of the field stars shows $K_s$ magnitude to be correlated with $J-K$ colour. The offset in $K$ magnitude of a star $i$ from the mean relation is given by
$$
\delta K_{fi} = K_i - 11.700 + 2.808((J-K)_i-0.461),
$$
with an approximately gaussian intrinsic scatter $\Sigma_{fK}=1.38$ mag.

\begin{table*}
\caption{Cross-identifications, 2MASS photometry and proper motions of candidate rotational variables with membership probabilities greater than 0.5. The 1SWASP identifier encodes the J2000.0 coordinates of the object. $\theta$ denotes the angular distance of the star from the cluster centre. 
The spectral types are estimated from 2MASS photometry using the calibration of \citet{blackwell1994}.
}
\label{tab:starparms}
\begin{tabular}{cccccccccc}
\hline\\
1SWASP	&	USNO B-1.0	&	Cl Melotte 111	&	Spectral	&	$J-K$	&	$K_s$	&	$\theta$	&	$\mu_\alpha$	&	$\mu_\delta$	& Membership	\\
	&		&		&	type	&		&		&	(deg)	&	(mas/yr)	&	(mas/yr)	& probability \\
\hline\\
J114837.70+281630.5 &  1182-0217523 & New member & G8 &   0.430 &  8.585 &  7.976 & $-10.6\pm   0.7 $ & $ -7.6\pm	0.7 $ &  0.976465 \\
J115533.35+294341.7 &  1197-0194906 & New member & K3 &   0.635 &  9.057 &  7.117 & $ -7.9\pm   0.9 $ & $ -8.2\pm	1.1 $ &  0.993480 \\
J120052.24+271923.9 &  1173-0240221 & New member & K4 &   0.678 &  8.996 &  5.118 & $-11.9\pm   0.7 $ & $ -4.4\pm	0.7 $ &  0.997313 \\
J120222.86+225458.7 &  1129-0233236 & New member & K4 &   0.787 &  9.603 &  5.614 & $ -4.3\pm   1.3 $ & $  2.9\pm	1.1 $ &  0.818328 \\
J120757.72+253511.3 &  1155-0190903 & AV 189 & K3 &   0.628 &  8.906 &  3.410 & $-11.5\pm   1.2 $ & $ -7.7\pm	0.9 $ &  0.998392 \\
J120836.10+310609.9 &  1211-0194040 & New member & K7 &   0.765 &  9.402 &  6.002 & $-16.8\pm   1.6 $ & $ -4.1\pm	2.0 $ &  0.977817 \\
J121135.15+292244.5 &  1193-0195921 & New member & K4 &   0.596 &  8.979 &  4.219 & $-12.5\pm   0.8 $ & $ -9.9\pm	0.9 $ &  0.997246 \\
J121253.23+261501.3 &  1162-0201218 & AV 523 & K2 &   0.587 &  8.990 &  2.284 & $-11.0\pm   1.1 $ & $ -9.0\pm	0.7 $ &  0.991622 \\
J122115.62+260914.0 &  1161-0200668 & AV 1183 & K4 &   0.642 &  8.972 &  0.420 & $-13.1\pm   1.4 $ & $ -9.2\pm	1.0 $ &  0.998253 \\
J122347.22+231444.3 &  1132-0216856 & AV 1404 & K4 &   0.659 &  9.018 &  2.760 & $-12.5\pm   0.6 $ & $ -8.4\pm	0.6 $ &  0.999326 \\
J122601.31+342108.4 &  1243-0197448 & New member & K5 &   0.702 &  9.143 &  8.378 & $ -2.1\pm   1.8 $ & $ -5.4\pm	2.2 $ &  0.922972 \\
J122651.03+261601.8 &  1162-0204591 & AV 1660 &K4  &   0.699 &  9.156 &  0.905 & $-13.3\pm   1.3 $ & $ -4.4\pm	1.1 $ &  0.992839 \\
J122720.68+231947.4 &  1133-0208375 & AV 1693 & G8 &   0.461 &  8.451 &  2.847 & $-11.5\pm   0.7 $ & $ -8.8\pm	0.6 $ &  0.999931 \\
J122748.29+281139.8 &  1181-0230534 & AV 1737, T141 &	G8 &   0.386 &  8.050 &  2.441 & $-13.1\pm   0.6 $ & $ -8.7\pm	0.6 $ &  0.999997 \\
J122856.43+263257.4 &  1165-0203085 & AV 1826, T220 &	K1 &   0.547 &  8.661 &  1.441 & $-12.6\pm   0.7 $ & $ -9.2\pm	0.8 $ &  0.999831 \\
J123231.07+351952.2 &  1253-0200298 & New member	 & G0 &   0.321 &  8.086 &  9.552 & $-12.2\pm   1.4 $ & $-10.7\pm	1.0 $ &  0.999935 \\
J123320.01+222423.3 &  1124-0241610 & New member & G8 &   0.453 &  8.402 &  4.297 & $-12.6\pm   0.7 $ & $ -8.5\pm	0.7 $ &  0.999955 \\
J123342.12+255634.0 &  1159-0194918 & AV 2177 & K0 &   0.447 &  8.584 &  2.406 & $-10.7\pm   0.7 $ & $ -7.7\pm	0.7 $ &  0.995428 \\
J123454.29+272720.2 &  1174-0253286 & AV 2257 & G0 &   0.387 &  7.510 &  3.031 & $ -9.8\pm   1.2 $ & $ -8.9\pm	0.6 $ &  0.550545 \\
J123811.47+233322.2 &  1135-0193392 & New member & M0 &   0.813 &  9.963 &  4.226 & $-12.0\pm   5.8 $ & $ -8.6\pm	5.8 $ &  0.522927 \\
J123941.99+213458.0 &  1115-0215169 & New member & K5 &   0.689 &  8.799 &  5.839 & $ -6.6\pm   0.7 $ & $ -3.5\pm	1.7 $ &  0.558604 \\
J124235.14+410527.7 &  1310-0230997 & New member & K5 &   0.683 &  9.273 & 15.626 & $ -7.1\pm   1.5 $ & $ -1.4\pm	2.1 $ &  0.804599 \\
J124900.42+252135.6 &  1153-0197249 & New member & K5 &   0.681 &  9.069 &  5.894 & $-10.3\pm   0.7 $ & $ -6.8\pm	0.7 $ &  0.997940 \\
J124930.43+253211.1 &  1155-0197399 & New member? & K5 &   0.682 &  8.931 &  5.985 & $-15.3\pm   1.4 $ & $ -6.0\pm	2.1 $ &  0.982752 \\
J125001.70+210312.1 &  1110-0214013 & New member & K3 &   0.582 &  8.976 &  7.925 & $-13.4\pm   1.4 $ & $  8.1\pm	1.5 $ &  0.839267 \\
J125211.61+252224.5 &  1153-0197624 & New member &	F8 &   0.272 &  7.609 &  6.606 & $-12.8\pm   1.2 $ & $ -9.0\pm	1.0 $ &  1.000000 \\
J125314.67+240313.6 &  1140-0196476 & New member & K5 &   0.750 &  9.452 &  7.121 & $  5.3\pm   2.7 $ & $-17.1\pm	3.9 $ &  0.690814 \\
J125736.86+285844.7 &  1189-0206667 & New member	 &	G8 &   0.424 &  8.473 &  8.231 & $-12.5\pm   0.6 $ & $ -7.5\pm	0.7 $ &  0.999222 \\
J125927.75+194115.1 &  1096-0208690 & New member	 & K5 &   0.692 &  9.034 & 10.502 & $ -9.3\pm   5.3 $ & $ -9.7\pm	5.5 $ &  0.877254 \\
J130543.99+200321.4 &  1100-0213302 & New member	 & K2 &   0.605 &  8.922 & 11.481 & $  3.5\pm   0.6 $ & $ -6.4\pm	1.0 $ &  0.989022 \\
\hline\\
\end{tabular}
\end{table*}

We multiply the probability densities for the $K_s$ magnitude offset from the field distributions with the field proper-motion probability density from Eq.~\ref{eq:probdensf} above to obtain
\begin{equation}
g_f =\frac{f_f}{\sqrt{2\pi}\Sigma_{fK}}
\exp\left(-\frac{\delta K_{fi}^2}{\Sigma_{fK}^2}\right),
\end{equation}
and similarly for the cluster probability density we obtain
\begin{equation}
g_c =\frac{f_c}{\sqrt{2\pi}\Sigma_{cK}}\exp\left(-\frac{(K_i-K_{\rm isochrone})^2}{\Sigma_{cK}^2}\right).
\end{equation}
The combined membership probability for an individual star is  then 
$$
p=\frac{g_c}{g_c+g_f}.
$$

In Fig.~\ref{fig:rank_pmem} we plot the cluster membership probabilities of the 50 rotational variables that are the most likely cluster members, in descending order of rank. The properties of 
\textcolor{black}
{the 30 stars with membership probabilities greater than 0.5}
are listed in Table~\ref{tab:starparms}, together with cross-identifications with previous catalogues of cluster members. 

\begin{figure}
\includegraphics[width=7.5cm,angle=270]{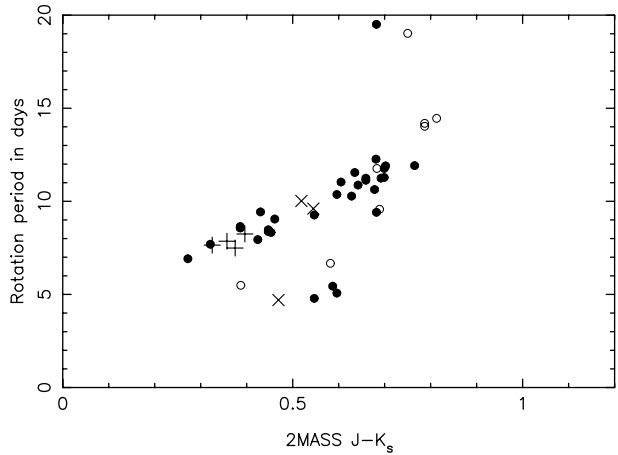}
\caption[]{Photometric rotation period as a function of 2MASS ($J-K$) colour for stars with significant (FAP$ < 10^{-2}$) quasi-sinusoidal variability and cluster membership probability greater than $0.5$, in the WASP fields surveyed in 2004 and 2007. Periods measured for candidates with cluster membership probabilities greater than 0.85 are denoted by filled circles; those with membership probabilities between 0.5 and 0.85 are denoted by open circles. 
Additional stars from \citet{radick90coma} and \citet{marilli97coma} are denoted by `+' and `x' symbols respectively.
}
\label{fig:col_per}
\end{figure}

\section{Period-colour relation}
\label{sect:percol}

The rotation periods of all stars with false-alarm probabilities less than 0.01 and membership probabilities greater than 0.5  are plotted as a function of $J-K$ colour in Fig.~\ref{fig:col_per}. Among these, the 
\textcolor{black} 
{periods}
of stars with cluster membership probabilities greater than 0.85 are denoted by solid symbols. They follow a tight period-colour relation. The properties of these stars are listed in Table~\ref{tab:starparms}, including cross-identifications with previous catalogues of cluster members. The rotational parameters of individual light curves that yielded significant detections in the 2004 and 2007 seasons are listed in Table~\ref{tab:fitparms}.

Several light curves yield periods that follow a second sequence in Fig.~\ref{fig:col_per}, with periods close to half those on the main relation. A similar secondary period-colour sequence was noted by \citet{hartman2008m37} in their recent study of the comparably-aged cluster M37. Several of them were recorded independently in two or more of the WASP cameras, and some in different seasons, and were found to jump between the two relations. The K4 dwarf 1SWASP J121135.15+292244.5, for example, yields $P=10.37$ d in 2004, and 5.07d in 2007. Similarly, the K1 dwarf J122856.43+263257.4 yielded periods of 4.79 d in 2004 and 9.26d in 2007. The K2 dwarf 1SWASP J121253.23+261501.2 yields 10.38d in 2004 and 5.44d in 2007, though only the latter measurement has FAP$ < 0.01$ and is listed here. 
 
\begin{figure*}
\includegraphics[width=11cm,angle=270]{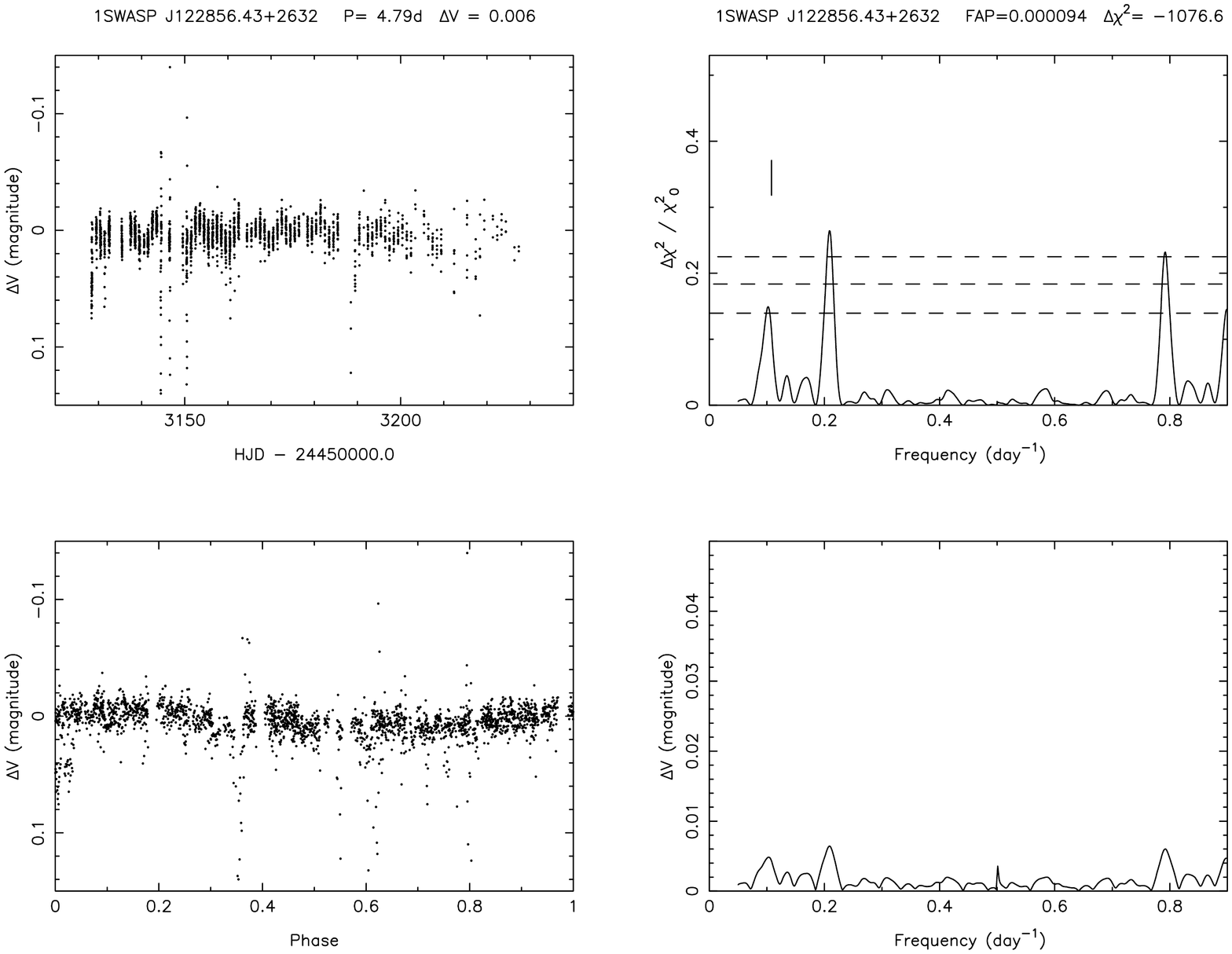}
\includegraphics[width=11cm,angle=270]{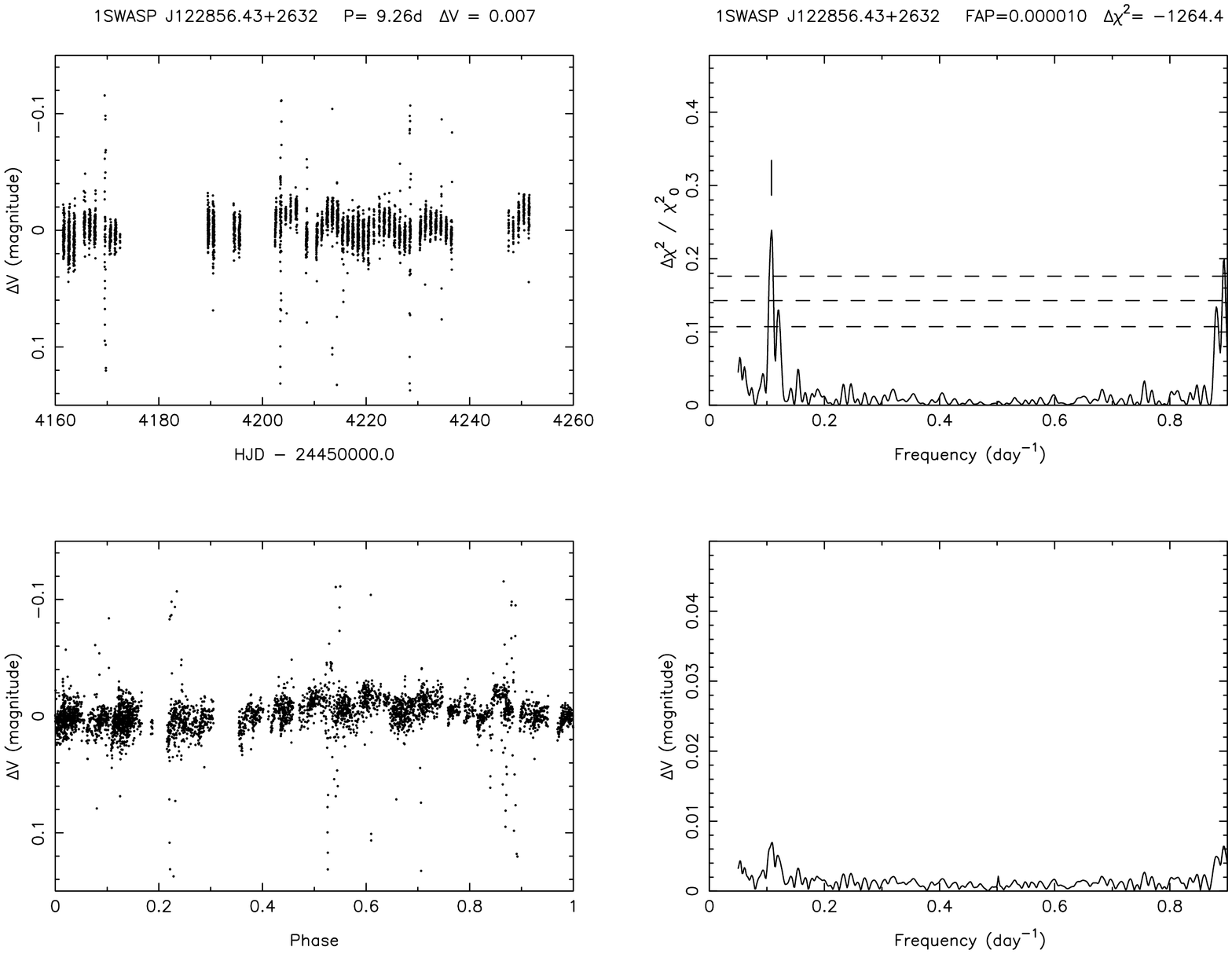}
\caption[]{Light-curves and periodograms for 1SWASP~J122856.43+263257.4 observed in 2004 (top) and 2007 (bottom). Each set of four panels shows the unphased light-curve at upper left. Occasional gaps in the data were caused by inclement weather. The periodogram of  $p=\Delta\chi^2/\chi^2_0$ versus frequency is at upper right, with horizontal dashed lines denoting thresholds of false-alarm probability 0.1, 0.01 and 0.001. The vertical bar indicates the true rotational frequency. The light curve phased on the best-fit period 
\textcolor{black}
{for the year concerned}
is shown at lower left, and the sine-wave amplitude periodogram at lower right. The 2004 data illustrate the frequency-doubling that occurs when major spot complexes are present on opposite stellar hemispheres, giving a double-humped light curve. }
\label{fig:J122856}
\end{figure*}

\begin{table*}
\caption{Sine-fitting periods for light curves of candidate rotational variables in the vicinity of Melotte 111, with cluster membership probabilities greater than 0.5. The year of observation is listed, with a designator incorporating the
RA and Dec of the field centre and a three-digit camera identifier for each light curve.  The number $\rmsub{N}{obs}$ of observations is listed, 
followed by the period of the peak in the periodogram yielding the greatest improvement in the $\chi^2$ statistic. Columns 6 and 7 give the 
amplitude of the fitted sinusoid and the false-alarm probability of the signal detection. 
Periods suspected of being half the true period appear in parentheses.}
\label{tab:fitparms}
\begin{tabular}{cccrrrrrr}
\hline\\
1SWASP	&	Season	&	Field\_cam	&	Nobs	&	Period	&	Amplitude	&	FAP	\\
	&		&		&		&	(days)	&	(mag)	&		\\
J114837.70+281630.5 & 2004 & SW1143+3126\_102 &  1138 &  9.43 & 0.005 & 0.000079   \\ 
J115533.35+294341.7 & 2004 & SW1143+3126\_102 &  1139 & 11.55 & 0.005 & 0.001371   \\ 
J120052.24+271923.9 & 2007 & SW1217+2350\_142 &  2841 & 10.63 & 0.010 & 0.000374   \\ 
J120222.86+225458.7 & 2004 & SW1217+2326\_104 &  1149 & 14.03 & 0.004 & 0.007144   \\ 
J120222.86+225458.7 & 2007 & SW1217+2350\_142 &  2824 & 14.19 & 0.008 & 0.001182   \\ 
J120757.72+253511.3 & 2007 & SW1217+2350\_142 &  2841 & 10.28 & 0.005 & 0.003890   \\ 
J120836.10+310609.9 & 2007 & SW1222+3000\_144 &  2960 & 11.92 & 0.007 & 0.002868   \\ 
J121135.15+292244.5 & 2004 & SW1216+3126\_103 &  1034 & 10.37 & 0.009 & 0.001038   \\ 
J121135.15+292244.5 & 2007 & SW1222+3000\_144 &  2959 &  (5.07) & 0.006 & 0.001124   \\ 
J121253.23+261501.3 & 2007 & SW1217+2350\_142 &  2840 &  (5.44) & 0.006 & 0.006021   \\ 
J122115.62+260914.0 & 2004 & SW1217+2326\_104 &  1147 & 10.88 & 0.011 & 0.000146   \\ 
J122347.22+231444.3 & 2004 & SW1217+2326\_104 &  1149 & 11.13 & 0.010 & 0.003327   \\ 
J122347.22+231444.3 & 2007 & SW1217+2350\_142 &  2841 & 11.24 & 0.009 & 0.000749   \\ 
J122601.31+342108.4 & 2004 & SW1216+3126\_103 &  1034 & 11.90 & 0.023 & 0.001428   \\ 
J122651.03+261601.8 & 2004 & SW1244+2427\_101 &  2269 & 11.77 & 0.005 & 0.000632   \\ 
J122651.03+261601.8 & 2007 & SW1222+3000\_144 &  2910 & 11.28 & 0.006 & 0.008509   \\ 
J122656.48+224054.7 & 2004 & SW1244+2427\_101 &  2265 &  (5.77) & 0.020 & 0.006039   \\ 
J122720.68+231947.4 & 2004 & SW1244+2427\_101 &  2183 &  9.05 & 0.006 & 0.000308   \\ 
J122748.29+281139.8 & 2007 & SW1222+3000\_144 &  2904 &  8.65 & 0.010 & 0.002253   \\ 
J122748.29+281139.8 & 2007 & SW1238+3135\_143 &  3255 &  8.57 & 0.012 & 0.000008   \\ 
J122748.29+281139.8 & 2007 & SW1242+2418\_141 &  3575 &  8.59 & 0.011 & 0.000008   \\ 
J122856.43+263257.4 & 2004 & SW1244+2427\_101 &  2268 &  (4.79) & 0.006 & 0.001287   \\ 
J122856.43+263257.4 & 2007 & SW1242+2418\_141 &  3833 &  9.26 & 0.007 & 0.006194   \\ 
J122942.15+283714.6 & 2007 & SW1222+3000\_144 &  2894 & 16.05 & 0.004 & 0.003018   \\ 
J123231.07+351952.2 & 2004 & SW1243+3126\_102 &  2259 &  7.69 & 0.008 & 0.000009   \\ 
J123320.01+222423.3 & 2007 & SW1242+2418\_141 &  3833 &  8.33 & 0.006 & 0.001505   \\ 
J123341.88+291401.7 & 2007 & SW1222+3000\_144 &  2790 & 16.88 & 0.010 & 0.006104   \\ 
J123342.12+255634.0 & 2004 & SW1217+2326\_104 &  1147 &  8.38 & 0.010 & 0.003695   \\ 
J123342.12+255634.0 & 2007 & SW1242+2418\_141 &  3831 &  8.47 & 0.009 & 0.000000   \\ 
J123354.22+270804.7 & 2007 & SW1222+3000\_144 &  2722 & 16.54 & 0.008 & 0.006181   \\ 
J123454.29+272720.2 & 2004 & SW1216+3126\_103 &  1017 &  (5.49) & 0.013 & 0.004207   \\ 
J123811.47+233322.2 & 2007 & SW1242+2418\_141 &  3828 & 14.46 & 0.013 & 0.000240   \\ 
J123941.99+213458.0 & 2004 & SW1244+2427\_101 &  2264 &  9.57 & 0.004 & 0.007075   \\ 
J124235.14+410527.7 & 2007 & SW1241+3924\_148 &  3548 & 11.76 & 0.003 & 0.003510   \\ 
J124309.53+244705.2 & 2007 & SW1242+2418\_141 &  3830 &  2.77 & 0.015 & 0.000027   \\ 
J124523.35+425104.4 & 2007 & SW1241+3924\_148 &  3542 & 12.05 & 0.030 & 0.002383   \\ 
J124900.42+252135.6 & 2007 & SW1242+2418\_141 &  3822 & 12.27 & 0.009 & 0.000001   \\ 
J124930.43+253211.1 & 2004 & SW1244+2427\_101 &  2264 & 19.51 & 0.017 & 0.000018   \\ 
J124930.43+253211.1 & 2007 & SW1242+2418\_141 &  3820 &  (9.41) & 0.019 & 0.000082   \\ 
J125001.70+210312.1 & 2004 & SW1244+2427\_101 &  2267 &  (6.67) & 0.003 & 0.003746   \\ 
J125211.61+252224.5 & 2007 & SW1242+2418\_141 &  3815 &  6.92 & 0.005 & 0.007111   \\ 
J125314.67+240313.6 & 2004 & SW1244+2427\_101 &  2267 & 19.02 & 0.027 & 0.000555   \\ 
J125419.08+324935.1 & 2004 & SW1243+3126\_102 &  2291 & 15.69 & 0.003 & 0.002814   \\ 
J125736.86+285844.7 & 2004 & SW1243+3126\_102 &  2279 &  7.94 & 0.007 & 0.003388   \\ 
J125927.75+194115.1 & 2007 & SW1252+1735\_147 &  3788 & 11.24 & 0.025 & 0.000021   \\ 
J130543.99+200321.4 & 2007 & SW1252+1735\_147 &  3625 & 11.04 & 0.003 & 0.002140   \\ 
\hline\\
\end{tabular}
\end{table*}

\begin{figure}
\includegraphics[width=7.5cm,angle=270]{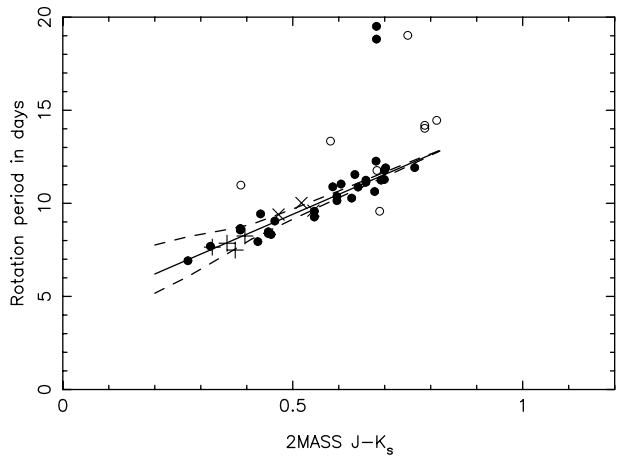}
\caption[]{As for Fig.~\ref{fig:col_per}, with the periods of the light curves on the lower branch having been doubled.  The solid line is a least-squares linear fit to the period-colour relation, and the dashed lines delineate the scatter about the mean relation expected from cyclic changes in spot latitude on differentially-rotating stars.
}
\label{fig:col_per_ext}
\end{figure}

These stars almost certainly rotate with periods near the main relation. The reason for the intermittent frequency doubling is apparent from Fig.~\ref{fig:J122856}. The 2004 light curve yields a period half that of the 2007 light curve, but the longer period is clearly present in both seasons. Evidently the apparent doubling of these stars' rotational frequencies at some epochs of observation arises from two dominant starspot groups on opposing stellar hemispheres giving a double-humped light curve. We conclude that the true periods of the light curves on the lower sequence are twice those derived from sine fitting. After correcting for frequency-doubling, the period-colour relation is as shown in Fig.~\ref{fig:col_per_ext}. 

Among the stars with membership probabilities greater than 0.85, only one outlier is seen with a period that departs significantly from the main relation. This star, 1SWASP J124930.43+253211.1 ($J-K=0.682$), exhibited a  clear period of 19.51d in 2004, with a weak secondary minimum. In 2007 the period appears to be 9.41 d, but the minima alternate in depth indicating the true period to be 18.82d. \citet{meibom2009m35} noted the presence of several similarly anomalous slow rotators in their rotation study of M35, and proposed that partial or complete tidal synchronisation in stellar binary systems could be responsible. Radial-velocity observations of this star would be desirable to check whether it really belongs to the cluster, and to test for evidence of binarity.

Half of the 23 rotational variables with membership probabilities greater than 0.85 lie within 4.2 degrees of the cluster centre, while the other half lie in the periphery of the cluster, between 4 and 11.5 degrees from the cluster centre.  Among them, 14 appear to be previously uncatalogued cluster members. Rotational variability thus appears to be a useful method for identifying new members of the cluster and its extended halo. 

In Figs.~\ref{fig:col_per} and \ref{fig:col_per_ext} we also show previously-published photometric periods for Melotte 111 stars T65, T76, T85 and T132 \citep{radick90coma} and T203, T213 and T221 \citep{marilli97coma}, with their 2MASS colours. 

The photometric periods found by \citet{marilli97coma} for T203 and T221 lie close to the SuperWASP period-colour relation. The 4.7-day period they found for T213 is very close to half the period predicted by Eq.~\ref{eq:col_per}. The sine-fitting method reveals a 9.51-day modulation in the WASP light curve of this star, but the false-alarm probability was below the threshold value 0.01 that would qualify it as a secure detection. We adopt a period of 9.4 days for this object, twice that reported by Marilli et al.

Together with the four Radick stars and the three Marilli stars, but excluding the anomalous slow rotator 1SWASP J124930.43+253211.1, the WASP light curves for stars with high cluster-membership probabilities and low false-alarm probabilities lie along a well-defined, almost linear period-colour relation. A linear least-squares fit to the period-colour relation of these stars yields
\begin{equation}
P = 9.71 + 10.68(J-K-0.528).
\label{eq:col_per}
\end{equation}

The RMS scatter of the observed periods about the fitted period-colour relation is 0.19 days - just 2 percent of the mean period -- and is not improved by the inclusion of quadratic or higher terms. This scatter is so small that much of it is attributable to 
\textcolor{black}
{year-to-year changes in the photometric modulation periods of} 
the cluster stars themselves. There are seven instances in which periods were measured for the same star in both 2004 and 2007. These season-to-season period differences are 0.09, 0.11, 0.16, 0.23, 0.32 and 0.49 days. The season-to-season differences in the period determinations for a given star average 0.23 days, comparable to the scatter about the mean relation. \citet{radick95hyades} reported finding very similar season-to-season period variations of between 2 and 8 percent in their 12-year study of rotation in the Hyades. 

There are two possible explanations for this trend: latitudinal differential rotation and core-envelope decoupling, both of whose effects will be superimposed on a spread at all spectral types resulting from a range of disc-locking lifetimes.  

Differential rotation on the stars themselves will lead to secular changes in the modulation period, if activity cycles are present which cause the main spot belts to drift in latitude. 
Doppler imaging and line-profile studies of differential rotation in young main-sequence stars show a strong decrease in latitudinal differential rotation along the main sequence. F and G dwarfs exhibit stronger shear than mid-to-late K dwarfs which rotate almost as solid bodies (\citealt{barnes2005diffrot}; \citealt{reiners2006diffrot}). \citet{cameron2007diffrot} gives the empirical relation for the equator-to-pole difference in rotational frequency
\begin{equation}
\Delta\Omega = 0.053 (T_{\mbox{eff}}/5130)^{8.6} {\rm rad\ d}^{-1}.
\end{equation}

We expect the resulting scatter in rotation frequency to be no more than half this amount, if the main active belts migrate across a similar range of latitudes as the solar butterfly diagram occupies on the Sun. The corresponding range of rotation periods at each colour is shown in Fig.~\ref{fig:col_per_ext}. The dispersions about the main period-colour reported here and by \citet{hartman2008m37} in M37 are comparable to the scatter expected from differential rotation for the F and G stars. The K stars also show season-to-season period changes that are most easily explained by a degree of differential rotation somewhat greater than Doppler imaging studies have found among more rapidly-rotating stars of similar mass. At all colours we find the season-to-season period changes to contribute a significant fraction, though perhaps not all, of the overall scatter about the mean period-colour relation. 

Additional residual scatter is expected at all spectral types if the early spin rates of stars are regulated by torque balance with accretion discs having a range of lifetimes. Such ``disc-locking'' models are reasonably successful at explaining the distribution of stellar spin rates on the zero-age main sequence (\citealt{cameron95clus}; \citealt{keppens95rot}; \citealt{krishnamurthi97rot}; \citealt{allain98rot}). 

Once the star decouples from the disc, angular momentum is lost through the stellar wind alone, but the spindown behaviour of the stellar surface depends critically on the degree of coupling between the star's radiative interior and convective envelope.  \citet{stauffer87pleiades} sought to explain the rapid spindown of G dwarfs between the ages of the 50 Myr-old $\alpha$ Per cluster and the 70 Myr-old Pleiades by suggesting that magnetic braking early in a star's life may spin down only the outer convective layers at first, leaving a decoupled reservoir of angular momentum in the stellar radiative core. If the coupling timescale on which the stored angular momentum is transported from the core back into the envelope is longer than a few tens of Myr, outward angular momentum transport will spin up the envelope at intermediate ages. This will counter the effects of magnetic braking and delay full convergence to a single period-colour relation. The effect would be lessened in K dwarfs, whose radiative interiors comprise a smaller fraction of the total stellar moment of inertia. 

\begin{figure}
\includegraphics[width=7.5cm,angle=270]{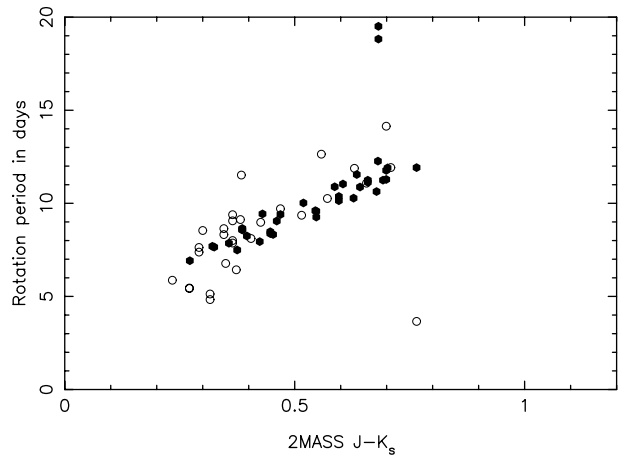}
\caption[]{Period-colour relations for Coma Berenices (filled circles) and the Hyades (open circles). The Hyades $J-K$ colours are taken from 2MASS, and the Hyades rotation periods from \citet{radick87hyades}.
}
\label{fig:col_per_hya}
\end{figure}

Several authors have modelled and discussed the implications of core-envelope decoupling for cluster period distributions at various ages (\citealt{li93rot}; \citealt{cameron94rot}; \citealt{krishnamurthi97rot}; \citealt{allain98rot}; \citealt{wolff2004rot}; \citealt{soderblom2001m34}). The models in these papers indicate that the weakest core-envelope coupling that can ensure the observed near solid-body rotation in the solar interior at an age of 4.6 Gyr would produce a dispersion by a factor two or more in the spin rates of solar-type stars at the age of the Hyades and Coma. The observed scatter among Coma stars with mid-F to mid-G spectral types 
is, however, only of order 2 percent. Since much of this is demonstrably caused by differential rotation, the evolutionary contribution to the scatter must be small. This  argues strongly for coupling timescales substantially shorter than the cluster age. 

\section{Gyrochronological ages of Coma and  Hyades}

The tight period-colour relation for the rotational variables found within 12 degrees of the core of the Coma Berenices cluster, sharing the proper motion and lying close to the cluster main sequence, contrasts strongly with the bimodal distribution of rotation periods seen in younger clusters. The absence of a population of rapid rotators in our survey is not, however, surprising because our faint limit extends only to the boundary between K and M spectral types. In the Hyades, rapid rotation is only seen at spectral types later than our faint limit. At earlier spectral types, the period-colour relation in Coma resembles the pattern of rotation rates observed in the Hyades by \citet{radick87hyades}, and subsequently updated by \citet{radick95hyades}. In Fig~\ref{fig:col_per_hya} we show the period-colour relations for both clusters, using 2MASS $J-K$ colours for the Hyades stars. 
\textcolor{black}
{All but five of the Radick et al. stars were included in the Hipparcos survey of the Hyades by \citet{perryman98hyades}, and were identified as cluster members from their parallaxes, proper motions and radial velocities.}

\textcolor{black}
{The relative rotational ages of the two clusters can be inferred from their respective period-colour relations using gyrochronology, assuming a simple rotational spindown model. We} follow a methodology similar to that of \citet{barnes2007gyro}, assuming the rotation period of a given star to be a separable function of age and colour. We divide the periods of stars in both clusters by the periods derived from their $J-K$ colours at the age of Coma using  eq.~\ref{eq:col_per}. Assuming that rotation period 
\textcolor{black}
{increases as} 
the inverse square root of age, the age of an individual star relative to 
\textcolor{black}
{the fiducial rotational age $t_{\rm Coma}$ }
of the Coma population is given by 
\begin{equation}
t=t_{\rm Coma}\left(\frac{P}{9.71 + 10.68(J-K-0.528)}\right)^2.
\label{eq:gyro_age}
\end{equation}

\begin{figure}
\includegraphics[width=7.5cm,angle=270]{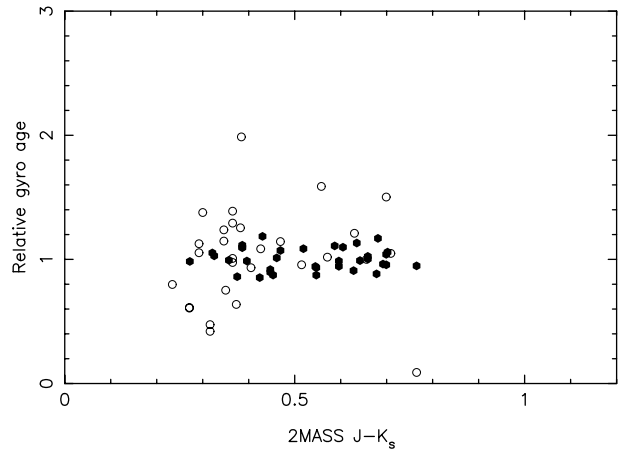}
\caption[]{Gyrochronological ages for individual light curves of stars in Coma Berenices (filled circles) and the Hyades (open circles), derived from Eq.~\ref{eq:gyro_age}.
}
\label{fig:gyro_age}
\end{figure}

\begin{figure}
\includegraphics[width=7.5cm,angle=270]{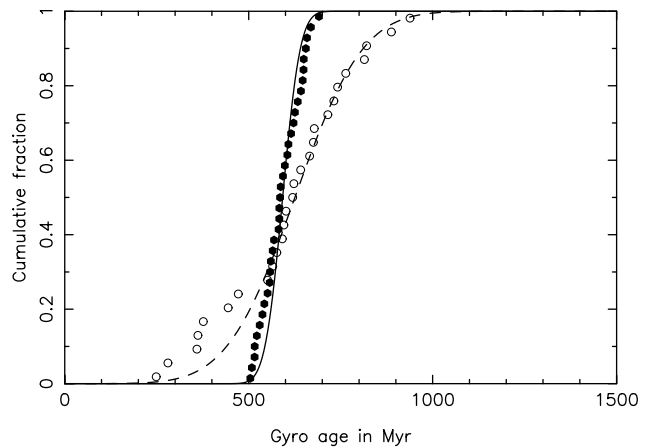}
\caption[]{Cumulative distributions of gyrochronological ages for individual light curves of stars in Coma Berenices (filled circles) and the Hyades (open circles), derived from Eq.~\ref{eq:gyro_age}. The mean age of the Coma Berenices stars is 591 Myr; that of the Hyades is 625 Myr.
}
\label{fig:agecompare}
\end{figure}

The relative gyrochronological ages of individual stars in the Hyades and Coma are shown as a function of $J-K$ in Fig.~\ref{fig:gyro_age}. 
The sample means for the relative ages of the Coma and Hyades stars are $1.002\pm 0.015$ and $1.058\pm 0.065$ respectively. These values exclude the lone Hyades rapid rotator with $J-K=0.77$, and the anomalous slow rotator discussed above at $J-K=0.682$. The difference between the gyrochronological ages of the two clusters is not statistically significant, owing to the high RMS scatter in the relative age determinations for the Hyades stars. The distribution of ages for individual Coma stars is much tighter, showing an RMS scatter of only 0.090 in relative age. If we adopt  625 Myr as the age of the Hyades cluster \citep{perryman98hyades}, the relative gyrochronological age of Coma is  found to be $590.7\pm 40.9$ Myr. 
The uncertainty is dominated by the scatter of the Hyades measurements, as seen from the cumulative distributions in Fig.~\ref{fig:agecompare}.  The mean gyrochronological age of the Coma sample is  thus $34 \pm 41$ Myr younger than that of the Hyades sample. This is in reasonable agreement with the recent comparative study of the isochrone ages of these clusters by \citet{king2005isoc}, who found Coma to be up to 100 Myr younger than the Hyades.
\textcolor{black}
{Adopting $t_{\rm Coma}=591$ Myr as the age of Coma, }
we derive the period-age-colour relation, 
\begin{equation}
P = \left(9.71 + 10.68(J-K-0.528)\right)\sqrt{t/591\mbox{ Myr}}.
\label{eq:col_per_age}
\end{equation}

\begin{table*}
\caption{J-K colours and spin periods for the Sun and old main-sequence stars with measured rotation periods and asteroseismological age determinations. The $J-K$ colours were determined using the effective temperatures and interpolating formulae of \citet{ramirez2005irfm}. The rotation periods are from \citet{donahue96} for the Sun, \citet{barnes2007gyro} for $\alpha$ Cen, and from \citet{stimets80} and \citet{noyes84} for 70 Oph A. The last row is based on the supposition that the true period of 70 Oph A is twice the measured period.The asteroseismological ages for $\alpha$ Cen and 70 Oph A are from Eggenberger et al. (2004, 2008). The gyro ages are estimated for two different values of the magnetic braking power-law index $b$.  }
\label{tab:gyrocal}
\begin{tabular}{lccccc}
\hline\\
Star                   & $J-K$ & Period              & Gyro Age & Gyro Age & Seismo          \\
                         &           & (days)              & $(b=0.5)$    & $(b=0.56)$  & Age (Gyr)       \\
\hline \\
Sun                   & 0.354 & $26.1\pm 2.0$ & $6.5\pm 1.0$ & $5.0\pm 0.7$ & 4.57                \\
$\alpha$ Cen A & 0.35  & $28.0\pm 3.0$  & $7.6\pm 1.7$ & $5.8\pm 1.1$ & $6.5\pm 0.3$  \\
$\alpha$ Cen B & 0.49  & $36.9\pm 1.8$  & $9.3\pm 0.9$ & $6.9\pm 0.6$ & $6.5\pm 0.3$  \\
70 Oph A          & 0.55   & $19.9\pm 0.5$  & $2.4\pm 0.1$ & $2.0\pm 0.1$ & $6.2\pm 1.0$  \\
70 Oph A          & 0.55   & $39.8\pm 1.0$  &  $9.5\pm 0.5$ & $7.0\pm 0.3$ & $6.2\pm 1.0$  \\
\hline\\
\end{tabular}
\end{table*}

Eq.~\ref{eq:col_per_age} yields a predicted sidereal solar photometric modulation period of 21.8 days for an assumed $(J-K)_\odot=0.354$ \citep{ramirez2005irfm} and a solar age $t_\odot=4.56$ Gyr. This is shorter than the observed 26.1-day sidereal period of sunspots in the mid-latitude active belt \citep{donahue96}. Conversely, the Sun's 26.1-day period yields an age of 6.5 Gyr using Eq.~\ref{eq:gyro_age} with $t_{\rm Coma}=591$ Myr. While an idealised braking law of this kind is convenient, it is unlikely to be appropriate for real stars whose moments of inertia evolve throughout their main-sequence lifetimes. A convenient  first-order correction may be a simple adjustment to the index $b$ of the braking  power law $P\propto t^b$, as \citet{barnes2007gyro} has proposed.

The major obstacle to fine-tuning the braking power-law index is the lack of old stars with reliably-determined ages and spin periods. 
A very few stars have already had their spin periods determined from the rotational modulation of their chromospheric Ca II $H\& K$ emission, and their ages determined from their solar-like $p$-mode oscillation patterns, namely $\alpha$ Cen A and B, and 70 Oph A. \citet{barnes2007gyro} found $\alpha$ Cen A and B to have comparable gyrochronological ages, from their 
respective spin periods of 28.0 and 36.9 days; \citet{eggenberger2004} derived an asteroseismological age of $6.5\pm 0.3$ Gyr for the $\alpha$ Cen system. The spin period of 70 Oph A was found by \citet{stimets80} to be 20.1 days and by \citet{noyes84} to be 19.7 days from the same set of Mt Wilson Ca II $H\& K$ modulation data. \citet{eggenberger2008} determine an asteroseismological age of $6.2\pm 1.0$ Gyr for 70 Oph A.

In Table~\ref{tab:gyrocal} we show that the rotational ages of these stars are best reconciled with the asteroseismological ages by adjusting the braking power law index from $b=0.5$ to $b=0.56$. While neither value of the index $b$ gives an age for 70 Oph A that agrees with the asteroseismological age of 6.2 Gyr at a spin period of 20 days, we note that Ca II $H\& K$ light curves can also exhibit double-peaked modulation in some seasons. It is possible that the true spin period of 70 Oph A is close to 40 days; future analyses of rotational splitting in the low-order $p$-modes of 70 Oph A would provide a valuable independent check on this star's rotation rate. If the longer period is correct, the asteroseismological and gyro ages can be reconciled using the same $b=0.56$ that fits the Sun and both components of $\alpha$ Cen. The best-fitting period-colour-age relation is then
\textcolor{black}
{
\begin{equation}
t=591\left(\frac{P}{9.30 + 10.39(J-K-0.504)}\right)^{1/0.56}\mbox{ Myr}.
\label{eq:gyro_age_tweaked}
\end{equation}
}
A departure of the braking power-law index from the $b=0.5$ of the Skumanich relation also seems to be needed in the comparison between K dwarfs in M35 and the Hyades \citep{meibom2009m35}. \citet{hartman2008m37} also find that they cannot fit M37, Hyades and the Sun simultaneously with a braking law that yields an asymptotic $t^{-1/2}$ age dependence for the stellar spin rate. \citet{barnes2007gyro} and \citet{mamajek2008gyro} find similar departures from the Skumanich relation in their gyrochronological age calibrations for cluster and field stars. For the purposes of the present investigation, however, the ages of the Hyades and Coma are in any case so similar that such small changes in the braking index will have little effect on the ratio of the two clusters' ages.

\section{Conclusions}

We have shown that inverse variance-weighted sine-fitting searches yield reliable period determinations from SuperWASP data in stars as faint as $K_s=10$, with amplitudes of rotational variability as low as 0.003 mag and periods of order 10 days recovered reliably with generalised Lomb-Scargle false-alarm probabilities less than 0.01. This has enabled us to carry out the first comprehensive study of rotation among the F, G and K dwarfs in the Coma Berenices open cluster. We find a narrow, nearly linear relation between rotation period and colour from mid-F to late-K spectral types.

The scatter of the individual stellar rotation rates about the mean period-colour relation is surprisingly small, as has previously been noted by \citet{radick87hyades} from their study of the Hyades, by \citet{hartman2008m37} and \citet{messina2008m37} from their more recent studies of M37, and by \citet{meibom2009m35} from their study of M35. Indeed, our period-colour relation for Coma has a markedly lower scatter about the mean relation than Radick et al. found for the Hyades. It is not clear whether the difference in scatter is intrinsic, or simply reflects the much denser sampling of the SuperWASP observations. Some of the rotation periods in the Radick et al. Hyades sample were derived from as few as 30 observations per star obtained over an interval of 5 months. The SuperWASP periods in Coma are derived from between 1000 and 3800 observations over a four-month period in each observing season. 

The  RMS scatter about the mean period-colour relation in Coma is less than 5 percent. A substantial fraction of this scatter appears to be attributable to season-to-season changes in period, caused by differential rotation and secular changes in active-region latitudes. The remaining few-percent residuals about the fitted period-colour relation are consistent with the not-quite-complete convergence in spin rates expected from rotational models at this age. We do not see the large scatter expected from models with core-envelope coupling timescales longer than about 100 Myr. Even solid-body spindown models leave a residual scatter in rotation rates at the age of Coma and the Hyades. This residual scatter depends somewhat on the detailed form of the braking law among the fastest rotators at early ages, but is seldom less than about 10 percent in the models cited above. 

Although the stars in these clusters have not entirely forgotten their original spin rates, the degree of convergence is good enough to yield reasonably accurate gyrochronological age estimates. The scatter in the period-colour relation propagates into the distribution of ages for individual cluster stars derived from a linear period-colour relation assuming a simple $t^{-1/2}$ spindown law \citep{skumanich72}. The most remarkable feature of Fig.~\ref{fig:agecompare}  is the sharpness of the age distribution inferred for Coma. The RMS scatter in the gyrochronological ages for the Coma stars is only just over 9 percent of the mean age derived for the ensemble. This high degree of internal consistency is echoed in M37, M35 and the Hyades, as other authors have noted previously. 

We have used $J-K$ as the colour index of choice for defining a period-colour-age relation based on the Coma data, partly because of the ready availability of well-determined $JHK$ colours in the 2MASS point-source catalogue for stars in the SuperWASP magnitude range, and partly because $J-K$ is less subject to reddening and metallicity effects than the more widely-used $B-V$
\textcolor{black}
{\citep{houdashelt2000}. }
Differences in metallicity from cluster to cluster may affect the $B-V$ colour at a given mass, and the period to which stars of a given mass converge at a given age. We raise the latter possibility because the angular momentum loss rate in a thermally-driven wind depends on the wind temperature as well as the total open magnetic flux threading the wind (\citealt{mestel87braking}; \citealt{cameron94rot}).  A dependence of wind temperature on metallicity  would lead to systematic differences in the asymptotic rotation rate for stars of a given mass and age in clusters with different abundances.

Despite these caveats, the relative ages derived here for Coma and the Hyades are in close agreement with the difference in their isochrone ages, 
\textcolor{black}
{even though, according to \citet{cayrel97abund}, the metallicity of Coma is sub-solar by about 0.05 dex, while the Hyades is overabundant by about +0.15 dex.} 
Our results thus support the contention of \citet{barnes2007gyro} that a properly-calibrated stellar rotational clock can provide a viable means of age determination in F, G and K stars that rotate more slowly than their  counterparts of the same colour  in the Hyades, Coma Berenices and M37 clusters. 

Some fine-tuning will still be needed to obtain gyrochronological ages of the best possible precision for main-sequence F, G and K stars. We contend that asteroseismology offers a promising way forward, providing precise measurements of both stellar ages and spin periods in mature main-sequence stars. Although such analysis is exacting and only feasible for very bright stars, it will provide valuable gyrochronological calibration standards at later ages. Gyrochronological ages can then be determined for much larger numbers of fainter stars, in a manner analogous to the cosmological distance ladder.

\section*{Acknowledgments}

The WASP Consortium consists of representatives from the Universities of
Keele, Leicester, The Open University,
Queens University Belfast and St Andrews, along with the Isaac Newton Group (La
Palma) and the Instituto de Astrofisica de Canarias (Tenerife). The SuperWASP
and WASP-S Cameras were constructed and operated with funds made available from
Consortium Universities and PPARC/STFC. This publication makes use of data products
from the Two Micron All Sky Survey,  which is a joint project of the University
of Massachusetts and the Infrared Processing and  Analysis Center/California
Institute of Technology, funded by the National Aeronautics and  Space
Administration and the National Science Foundation.This research has made use
of  the VizieR catalogue access tool, CDS, Strasbourg, France. 
\textcolor{black}
{We also thank the anonymous referee for insightful comments that led to substantial
improvements in our methodology.}

\bibliographystyle{mn2e}

\bsp

\label{lastpage}

\end{document}